# Inventions on
# Displaying and Resizing Windows
## A TRIZ based analysis

**Umakant Mishra**

Bangalore, India

http://umakantm.blogspot.in

**Contents**

## 1. Introduction

Windows are used quite frequently in a GUI environment. The greatest advantage of using windows is that each window creates a virtual screen space. Hence, although the physical screen space is limited to a few inches, use of windows can create unlimited screen space to display innumerable items.

The use of windows facilitates the user to open and interact with multiple programs or documents simultaneously in different windows. Sometimes a single program may also open multiple windows to display various items. The user can resize the windows and move their location time to time as desired.

However, there are several concerns of a window relating to its size, appearance, positioning, color, visibility, resizability etc. The following are some of the issues faced while using windows.

- ⇒ There is a concern regarding the size of windows, i.e., whether the window size should be allowed to be bigger than the physical screen size. If the window size cannot be bigger than screen size, then it cannot display larger items. But if the windows are expanded bigger than the screen size, then its borders may go beyond the screen and become inaccessible.

- ⇒ If we limit the size of the windows then there are issues like, what should be the maximum size of windows, how do the user know that the window cannot expand further, whether the size of one window should affect the size of other windows?

- ⇒ Resizing windows should not go beyond a minimum size, as otherwise, the user cannot know the content of the window.

- ⇒ Conventionally the user has to move the mouse pointer to the corner of the window to drag or resize the window. This is felt to be a stressful process when repeated again and again.

- ⇒ Holding the mouse pointer exactly on the window border or corner is felt to be difficult especially if you have a shaky hand, a fast pointer speed or a narrow border. In such cases there is possibility of clicking the mouse when the pointer is moved out of the window, which can cause unpredictable results.

- ⇒ When there are multiple windows, some windows partially or fully block other windows and hinder their visibility.

- ⇒ When a window is completely covered by other windows, it is difficult to select that window, as no part of the window is visible for the user to click and select.

- ⇒ Sometimes the screen space is occupied by unwanted windows. The user may leave the windows unclosed even after their requirements are over.

- ⇒ When the resolution is changed, the size of the window and contents of the window needs to be resized. If the resolution is too low, the size of the windows grow bigger leading to dislocation of windows and their contents. This may lead to inaccessibility of some items because of their positions beyond the screen area.

## 2. Inventions on displaying and resizing windows

### 2.1 Signaling the user attempting to resize an window beyond the limit (US Patent 5956032)

**Background problem**

In a windows environment there are several windows displayed on the screen for one or multiple applications. The user often resizes the windows to display them in most effective way within the limited screen space. There is a problem involved in this matter. When the user wants to drag the border of the window beyond the maximum limits, the border does not move further, but the interface does not show any signal indications. This creates confusion in the user as why the border does not move further. There is a need to inform the user that he is trying an impossible action.

**Solution provided by the invention**

Andrea Argiolas disclosed a method of resizing windows (Patent 5956032, Assigned by IBM, Sep 99), which visually indicates to the user that the window cannot be further extended. A graphical icon provides an immediate and perceivable feedback of the forbidden attempt.

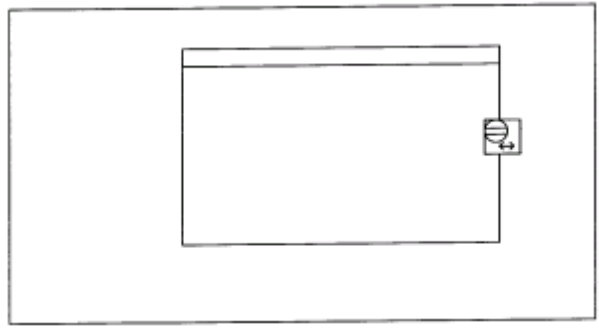

**TRIZ based analysis**

The invention provides a visual feedback to the user to communicate that the border is not further expandable (Principle-23: Feedback).

### 2.2 Method for managing simultaneous display of multiple windows in a graphical user interface (US Patent 6025841)

**Background problem**

In a typical windows environment, there are several windows displayed on the screen. The user clicks on the visible part of a window to bring it to the front. Thus the user has to click on different windows in order to see their contents.

There is a problem in this mechanism. When the size of the window is large, it covers other small windows. If a window is totally hidden behind another window then it becomes difficult to select that window, as the user cannot see any part of it to click. There is a need to display multiple windows in a more stable and persistent fashion so that the important windows do not get covered by other windows.

**Solution provided by the invention**

Finkelstein et al. found a method of managing simultaneous display of multiple windows (Patent 6025841, Assigned by Microsoft, Feb 2000). The invention provides a continuous, automatic adjustment to window size, and position of a selected top most window. The invention provides three different methods to keep the target window from obscuring the content of an underlying window.

1. Move away- when the window is selected to be on the top, it is moved away to a location where there is more free screen.
2. Disappear- the target window is disappeared momentarily and reappears once again when the action (say dragging an object) is completed.
3. Reduce- the system reduces the size of the target window to smaller, which is less obscuring.

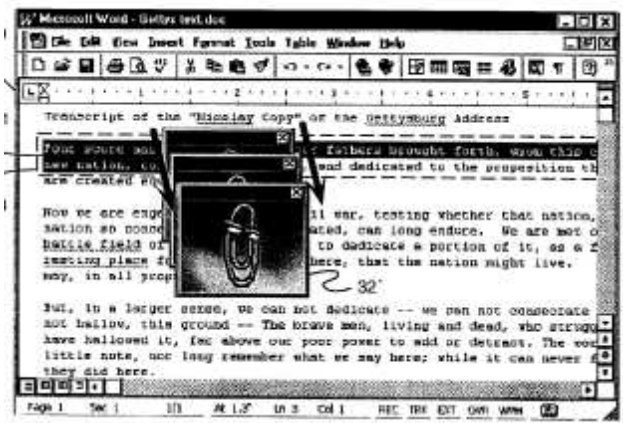

**TRIZ based analysis**

When the target window is hiding other useful contents of the screen (at the input of a trigger through mouse or keyboard) the system calculates a new location for the target window and redraws the window in that position (Principle-15: Dynamize).

The system displays the selected window typically in a higher z-order so that it is displayed on the top (Principle-17: Another dimension).

The windows are adjusted automatically to display the useful contents during operation based on the interaction of the user, without the user explicitly adjusting window size or position etc. (Principle-25: Self service).

The system uses three different methods to keep a target window from obscuring the contents of an underlying window, viz., Move away (Principle-15: Dynamize), Disappear (Principle-34: Discard and recover), and Reduce (Principle-35: Change parameter).

**2.3 New method of managing windows in GUI (US Patent 6181338)**

**Background problem**

The concept of "window" is an important feature of a Graphical User Interface. The user can open and interact with multiple programs or documents simultaneously in different windows. The user can resize the windows and move their locations from time to time depending upon the need.

According to the current convention, the user has to move the mouse pointer to the corner of the window to drag or resize the window. This is a difficult process and it becomes very stressful and non-productive when repeated again and again. There is a need for a more efficient method to move and resize windows more conveniently.

**Solution provided by the invention**

Mark Brodhun disclosed a new method of managing and controlling the size and location of windows in a GUI based computer system (Patent 6181338, assigned by IBM, Jan 2001). According to the invention, a user can quickly and easily relocate and resize a window without unnecessary movement of cursor. The user can invoke the window control mechanism by using both the buttons a two-button mouse. Alternatively the user can invoke the windows control mechanism by pressing a specific key combination on the keyboard.

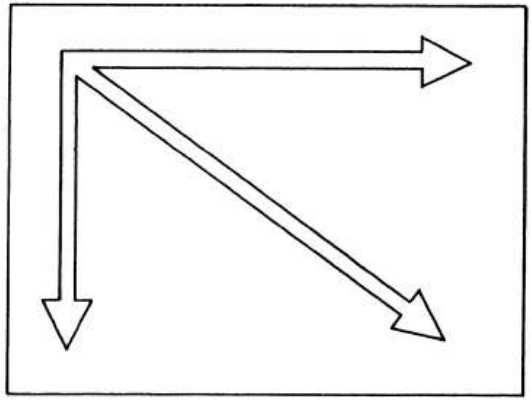

**TRIZ based analysis**

The invention uses a special key combination on the keyboard or mouse button instead of long and precise cursor movements to the window corners (Principle-28: Mechanics Substitution).

## 2.4 Method and system for automatically resizing and repositioning windows in response to changes in display (US Patent 6473102)

**Background problem**

Some display systems offer the user the ability to change the resolution of the display device. But increasing or decreasing the resolution results in decreasing and increasing the size of display items correspondingly, and often shifts the location of components from their original position. But there are some components, such as tool palettes and utility windows, which are not desirable to be shifted from their original positions.

Similarly, while switching the display from a full-sized monitor to an LCD screen, the positions and sizes of the objects on the display can change significantly, because of the operating parameters of the two display devices. This may even make some items unviewable because they are positioned outside the display area. It is necessary to provide a mechanism that is capable of maintaining windows and similar objects in an accessible condition even if there are changes in the display environment.

**Solution provided by the invention**

Rodden et. al, disclosed a method of automatically resizing and repositioning windows (Patent 6473102, assignee Apple Computers, Oct 2002) in response to changes in display environments or display parameters. According to the invention, the window position is recalculated and redrawn within the newly available display area. A minimum size is retained to display at least the minimum required information. For example, if the window is a utility window containing buttons, the minimum size of the window requires that at least one button be visible.

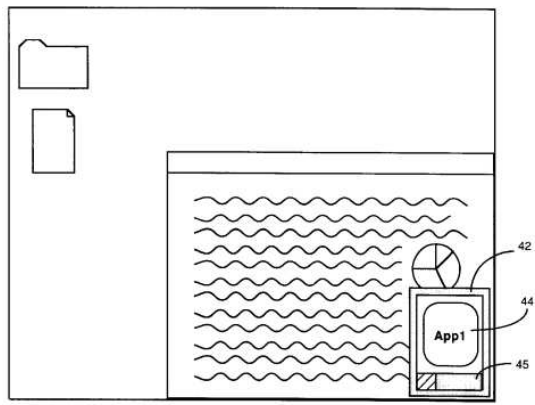

If the minimum size is not present in the available display area, the window is moved into the available area to attain the minimum size. Conversely, if there is more space after displaying the minimum information, the system decides whether to display additional integral components. After the window is displayed, the control elements, such as scroll buttons etc. are redrawn at the new size and position.

## TRIZ based analysis

The invention resizes and relocates the windows to be displayed in an altered display environment (Principle-15: Dynamize). However, although the components are resized, the positions of the windows remain in tact regardless of changes in the size or resolution of the display device (Principle-39: Calm).

If the space is not sufficient to display the window, the window shrinks to display the minimum required information. Conversely, if there is space can display more than the minimum required information, the system may determine to display additional integral components (Principle-16: Partial or excessive action).

## 2.5 Method, apparatus and computer program product for implementing graphical user interface window control (US Patent 6480207)

### Background problem

When resizing a window it is difficult and labor intensive to put the mouse pointer exactly on the window border or corner to get the double arrow icon for resizing. This is especially worse if you have a shaky hand, a fast pointer speed or a narrow border. It may so happen that you click the mouse button when the pointer is moved out of the window border and the operating system brings a new window to the foreground automatically. In that case you have to bring your window back into view again. There is a need for developing an easy and efficient method for resizing windows.

### Solution provided by the invention

Patent 6480207 (invented by Bates et al., assigned by IBM, issued in Nov 2002) provides a better window control mechanism, which overcomes the above difficulties of the prior art.

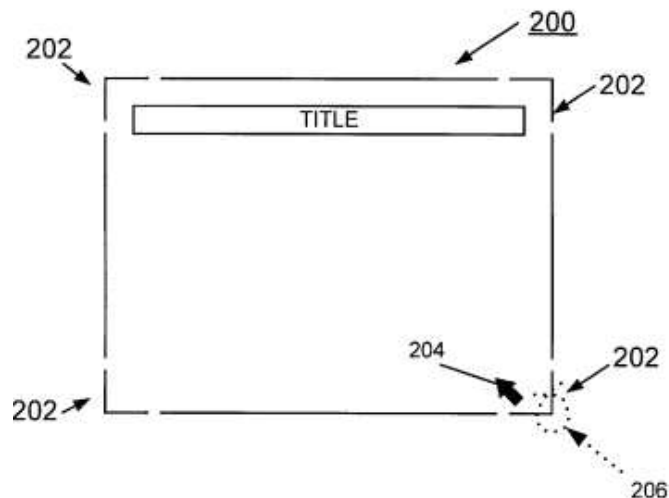

According to the invention a location queue stores the last N mouse locations within a defined maximum allowable time for a lasso to occur. The mouse selects the window border when the mouse pointer is moved around the selectable border portion within a time window.

When the user is drawing a cursive letter "e" or "o" straddled over the border the border is selected for resizing. This method is considered equivalent to throwing a rope around the border in a half hitch knot.

**TRIZ based analysis**

The selection of the border is not determined by clicking on the window corner but by moving the pointer around the corner point (Principle-28: Mechanics Substitution).

The invention intends to reduce the difficulty of resizing a window by a shaky hand. The invention wants the shaky hand to straddle over the window border so that the border is selected for resizing based on the time spent by the pointer around the border (Principle-8: Counterweight).

The invention not only considers the current location of the mouse (SPACE) but also considers previous N locations within a specified time (TIME) as a factor of selecting the border for resizing (Principle-17: Another dimension).

## 2.6 User interface and method for maximizing the information presented on a screen (US Patent 6512529)

**Background problem**

It is a conventional problem to display the contents of many windows in the limited screen space. There are different ways of displaying windows on a screen, such as, "tiled" where the contents are totally visible but the window size is small, "cascaded" where the window size is larger but the contents are partially overlayed by other windows. There is a need for a method to display multiple windows without obscuring the main window.

**Solution provided by the invention**

Janssen et al. invented a method (Patent 6512529, Assigned by Gallium Software, Jan 2003) for maximizing information presentation on a screen.

he invention prescribes 4 different display modes. In "normal mode", the contents of the windows are exposed. In "timed mode" the contents are displayed for a specified period of time. In "locked mode" the contents of the window are displayed in an opaque manner. In the "timed icon mode", the contents of the window are exposed for a specified period of time after which the window is automatically reduced to an icon.

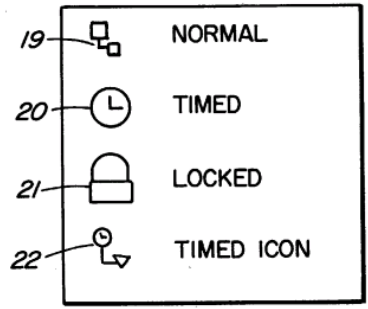

According to the invention the windows are made invisible so that the information on the background window is not obscured. The invention consists of a user interface that provides a rapid means of displaying and hiding information in invisible windows.

**TRIZ based analysis**

The invention accommodates more icons and windows, which are not displayed in the normal window, rather displayed in special modes and invisible trays. (Principle-17: Another dimension).

In "timed mode" the window returns to its invisible state after a specified period of time (Principle-32: Color change).

The user can change the status of any window as "normal", "timed", "locked" or "timed icon" to change the pattern of their visibility (Principle-15: Dynamize, Principle-35: Parameter change).

## 3. Summary

It is interesting to observe various solutions disclosed by different inventions to overcome different problems relating to displaying and resizing of windows. Some inventions disclose methods of automatic resizing and relocating of windows, some inventions display alternative modes of displaying windows to accommodate within the limited display area.

The difficulty of holding the window border (for resizing) can be overcome by combining both spatial (position of mouse pointer) and temporal (duration of the pointer around the vicinity of the window border) methods to select the window border for resizing.

In order to organize the windows within the available screen space, the status of some windows can be changed to minimized (or cascaded), some windows can be made invisible (say after being displayed for a specified period of time), some windows may disappear momentarily while the user executes an action and reappears once again after the action is completed, and some important windows may be locked so that other windows cannot obscure it.